\documentstyle[12pt,lathuile]{article}
\def\Journal#1#2#3#4{{#1} {\bf #2}, #3 (#4)}

\def\NPB{{ Nucl. Phys.} B}
\def\PLB{{ Phys. Lett.}  B}
\def\PRL{ Phys. Rev. Lett.}
\def\PRD{{ Phys. Rev.} D}

\def\simlt{\mathrel{\lower2.5pt\vbox{\lineskip=0pt\baselineskip=0pt
           \hbox{$<$}\hbox{$\sim$}}}}
\def\simgt{\mathrel{\lower2.5pt\vbox{\lineskip=0pt\baselineskip=0pt
           \hbox{$>$}\hbox{$\sim$}}}}
\def\A{{\cal A}}

\def\L{{\cal L}}

      \let\G=\Gamma   \let\L=\Lambda

 \def\bd{\begin{document}} \def\ed{\end{document}}
\def\ds{\documentstyle} \let\fr=\frac \let\bl=\bigl \let\br=\bigr
\let\Br=\Bigr \let\Bl=\Bigl 
\let\bm=\bibitem
\let\na=\nabla
\let\pa=\partial \let\ov=\overline
\def\ie{{\it i.e.\ }} 
\def\tr{{\mbox{\rm tr}}}
\def\st{\scriptstyle}
\def\sst{\scriptscriptstyle}
\def\mco{\multicolumn}
\def\epp{\epsilon^{\prime}}
\def\vep{\varepsilon}
\def\ra{\rightarrow}
\def\ppg{\pi^+\pi^-\gamma}
\def\vp{{\bf p}}
\def\ko{K^0}
\def\kb{\bar{K^0}}
\def\al{\alpha}
\def\ab{\bar{\alpha}}
\def\be{\begin{equation}}
\def\ee{\end{equation}}
\def\bea{\begin{eqnarray}}
\def\eea{\end{eqnarray}}
\def\CPbar{\hbox{{\rm CP}\hskip-1.80em{/}}}
\begin{document}
\title{
\vspace*{-0.8cm}   
\begin{flushright}   
 \normalsize{ETH-TH/01-11\\      
CERN-TH/2001-204}\\ 
 \end{flushright}    
\vspace{1cm}
\textbf{ LIMITS ON THE SIZE OF EXTRA DIMENSIONS  AND THE STRING SCALE
}$^\star$
\author{{\bf I.~Antoniadis~$^{1,2}$  and  K.~Benakli~$^{1,3}$} \\
{\em $^1$ CERN Theory Division CH-1211, Gen\`eve 23, Switzerland } \\
{\em $^3$ Theoretical Physics, ETH Zurich, Switzerland}
}}
\maketitle
\thispagestyle{empty}
\vspace*{.5cm}
\baselineskip=14.5pt
\begin{abstract}
We give a brief summary of present bounds on the size of possible 
extra-dimensions as well as the string scale from collider experiments.
\end{abstract}
\vspace{7.cm} 

\hspace{-1.cm}\small{$^\star$ Based on talks given by I.A. at La Thuile2001
 and by K.B. at Moriond2001 .}

\hspace{-1.cm}\small{$^2$ On leave of absence from CPHT, 
Ecole Polytechnique, UMR du CNRS 7644}%
\newpage

\section{Introduction}

Although the standard model has been experimentally verified to an impressive
precision level, it remains unsatisfactory in some of its theoretical 
aspects. The major one concerns dealing with the quantum effects of gravity. 
The renormalization 
procedure which allows to extract finite predictions for processes involving
the  three other fundamental forces fails when gravitational interactions 
are taken into account. String theory stands here as  only known consistent 
framework to incorporate these effects. There, known  
fundamental  particles are ``point-like''only because the
experimental energies are too small to excite the string oscillation modes
so only the center of mass motion is perceived. In addition to these heavy
oscillation modes, strings have new degrees of freedom  that  often 
take the classical geometry description of 
propagation in extra dimensions and thus provide a compelling reason for the 
latter.  This raises important questions:
Is it possible that our world has more dimensions than  the one that we are
aware of? If so, why don't we see the other dimensions? Is there
 a way to detect them?. Answering these questions has attracted a lot of
 efforts  lately as it has
became clear that not only these extra-dimensions might be there, but also
they could be just at the border of the energy domain 
at reach to near future experiments. This is because, within our present
knowledge,  the only requirements for the sizes of 
 compactification and string  scales are to allow the correct magnitude 
for the strength of the gauge and gravitational  couplings without falling in
 the already experimentally excluded regions and recent investigations
 ~\cite{Ant} indicate that there are many string vacua 
that allow the new physics to be at reach of LHC.

\section{ Compactification of extra dimensions }

Suppose that space-time has $D$ extra dimensions compactified
on a  $D$-dimensional torus of volume $(2 \pi)^D  R_1 R_2 \cdots R_D$. 
The  
states propagating in this $(4+D)$-dimensional space are seen from the
 four-dimensional point of view as a having a (squared) mass 
(assuming periodicity of the wave functions along each compact direction):
\be
M^2_{KK}\equiv M^2_{\vec n} = m_0^2 +\frac {n_1^2}{R_1^2} +
\frac {n_2^2}{R_2^2}+ \cdots  +\frac {n_D^2}{R_D^2}\, ,
\label{KKdef}
\ee
with $m_0$ the four-dimensional mass and $n_i$ non-negative
integers.  The states with $\sum_i n_i \neq 0$ are called Kaluza-Klein
(KK) states. An important remark is that not all states can propagate in the 
whole space. Some might be confined in subspaces with no KK excitations in the
transverse directions.
The simplest example of 
such a situation appears in compactification on $S^1/Z_2$ orbifolds obtained by gauging the  $Z_2$ parity:
$y \rightarrow - y \,  {\rm mod} \,   2\pi R$. where $y
\in [-\pi R, \pi R]$ span the fifth coordinate. The spectrum of states has 
some intersting properties: (i) only states invariant under this $Z_2$ (which
acts also on the gauge quantum numbers) are 
kept while the others are  projected out; (ii) new (``twisted'') states, 
localized at the end points have to be included. They have quantum numbers 
and interactions that were not present in the unorbifolded original 
5-dimensional model. As they
can not propagate in the extra-dimension, they have no KK excitations; 
(iii) The even states can have non-derivative renormalizable  couplings to
localized states. For instance, the couplings of massive
KK excitations of even gauge bosons  to localized fermions are given by:
\bea
g_{{\vec n}} = {\sqrt{2}}\sum_{{\vec
n}} e^{-\ln {\delta} \sum_i\frac{n_i^2l_s^2}{2 R_i^2}} g_0
\label{coupling}
\eea
where $l_s \equiv M_s^{-1}$ is the string length and
 $\delta=16$ in this case of
$Z_2$ orbifolding. The ${\sqrt{2}}$ comes from the relative normalization of
$\cos(\frac {n_i y_i}{R_i})$ wave function with respect to the zero mode
while the exponential damping is  a result of tree-level string computations
\cite{AB}.

Another example is obtained with intersecting branes
(see Figure 1 and 2).
When the angle between the intersecting branes is $\pi/2$ 
the localized strings behave exactly as the $Z_2$ twisted states described 
above. The exponential form factor of the coupling of KK excitations can be
viewed as the fact that the branes intersection has a
 finite thickness. In fact the interaction of the KK
excitations of the  gauge fields ( on the big branes) 
$A^\mu (x,\vec y)=\sum_{{\vec n}}
\A^{\mu }_{{\vec n}} \exp{i\frac {n_i y_i}{R_i}}$ with the charge
density $j_\mu (x)$ associated to massless  localized fermions is described
by the effective Lagrangian \cite{ABL}: 
\bea 
\int d^4x \, \, \,  \, \sum_{{\vec
n}} e^{-\ln {\delta} \sum_i\frac{n_i^2l_s^2}{2 R_i^2}} \, \,
\, \, \, j_\mu (x) \, \A^{\mu }_{\vec n}(x)\, , \eea 
 which can be written after Fourier transform as 
\bea \int d^{4}y \,\int d^4x  \, \,
\, \,  (\frac{1}{l_s^2 2 \pi \ln {\delta}})^{2} e^{- \frac {{\vec
y}^2}{2 l_s^2 \ln {\delta}}}  \, j_\mu (x) \, A^\mu (x,\vec y)\, .
\label{brwidth}
\eea
from which we read that the localized fermions are felt 
 as forming a Gaussian distribution of charge $e^{-\frac {{\vec y}^2}{2 \sigma^2}}
 j_\mu (x)$  with a width $\sigma=\sqrt{\ln {\delta}}\, l_s \sim 1.66 \, l_s$.

\begin{figure}[t]
 \vspace{9.0cm}
\includegraphics{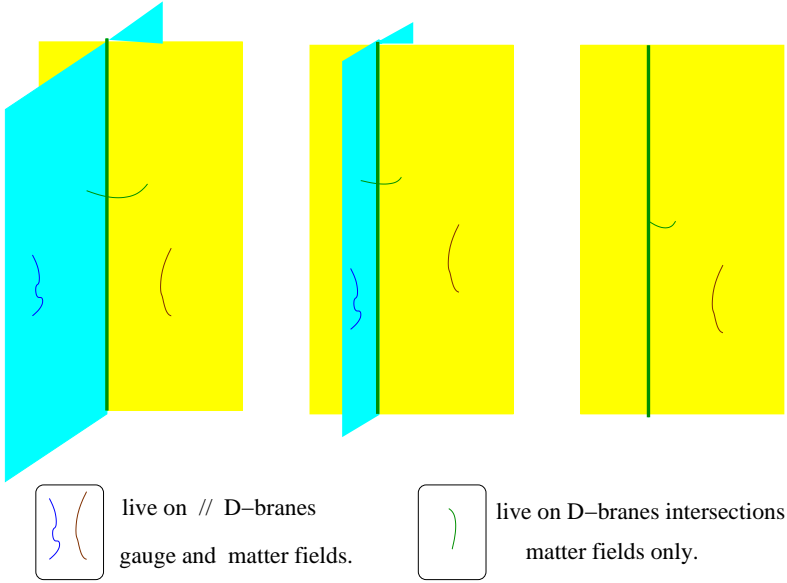}
 \caption{\it Zero modes of open strings 
 streched between two branes give rise to  matter localized at their
 intersection. One of the branes is shown ``losing'' one if its longitudinal
 dimensions as the size of the latter shrinks. The final result is  a small
 brane inside a bigger one
    \label{exfig1} }
\end{figure}
\begin{figure}[t]
 \vspace{9.0cm}
\includegraphics{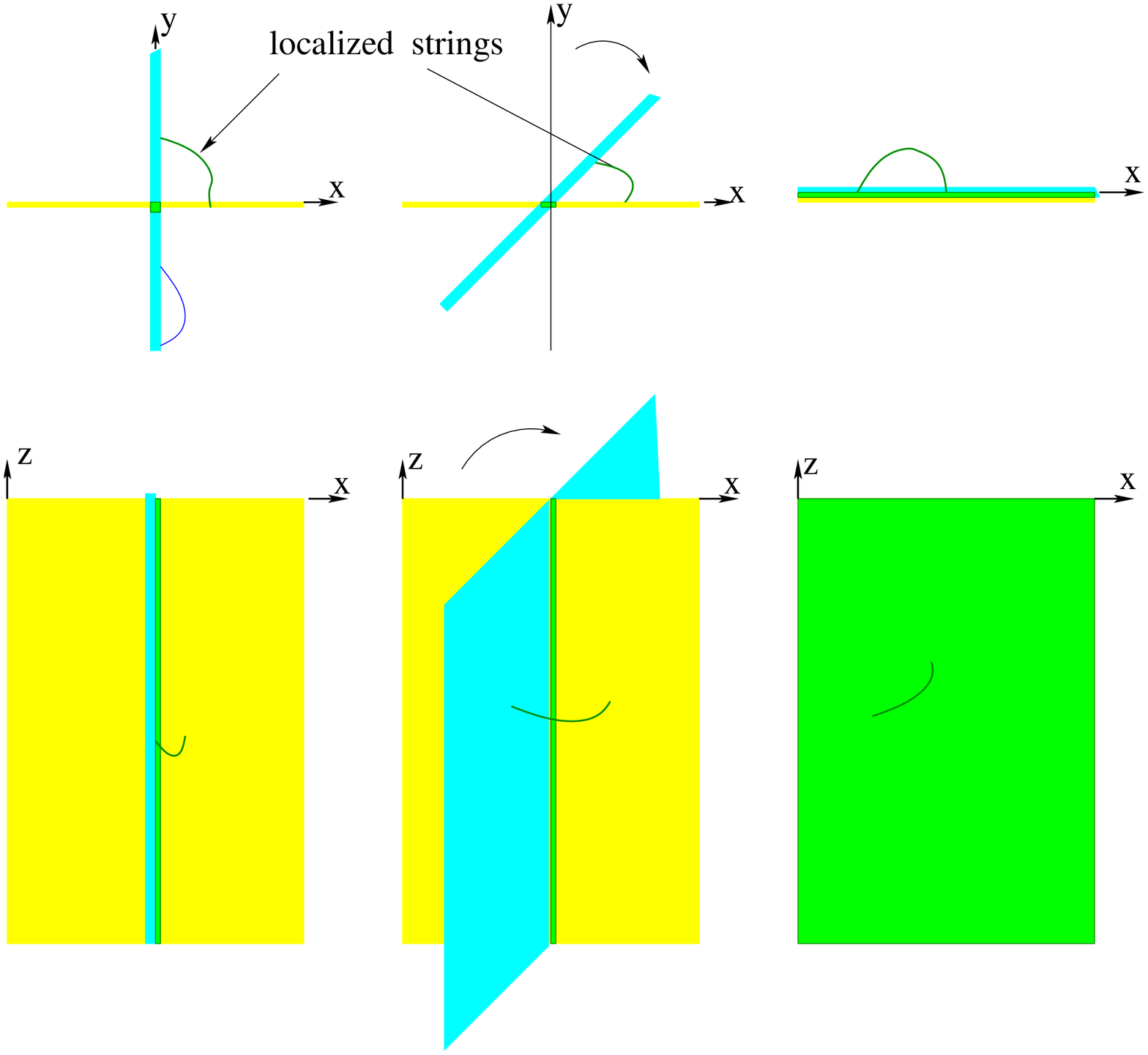}
 \caption{\it Rotating two branes from orthogonal position $\theta=\pi/2$ to
 parallel one $\theta=0$.
    \label{exfig2} }
\end{figure}

\section{Scattering of four localized fermions}

The total amplitudes for the scattering of  four fermions depend on the string coupling $g_s = g_{YM}^2$, the
string  scale $M_s \equiv 1/l_s$, the compactification radii $R_i$ and on
kinematical invariants that can be expressed in terms of the Mandelstam
variables $ s= -(k_1+k_2)^2\,  , \,t= -(k_2+k_3)^2 $ and $u=-(k_1+k_3)^2$. 
The result  can be decomposed as: 
\bea  \A = \A^{(0)}+ \A^{(KK)}+\A^{cont}_{w} +
\A^{cont}_{osc}\, ,
\label{defcon}
\eea where $\A^{(0)}$ is the contribution of the lightest  states (for example
from standard model fields), $\A^{(KK)}$ the one from  KK states of the form: \bea - \left[   {\bar
\psi}^{(1)} \gamma_{M} \psi^{(2)}  {\bar \psi}^{(4)} \gamma^{M} \psi^{(3)}
\right ] \, \,   \frac {g_s}{ l_s^{-4}\, \prod_{i=1}^{4} R_i} \sum_{m_i \in
{\bf Z}-\{ 0\}}  \frac {  \delta^{  - \sum_{i=1}^{4} \frac{m_i^2 l_s^2}{
R_i^2}}} { s - \sum_{i=1}^{4} \frac{m_i^2 }{ R_i^2}}\, , \nonumber \\
\label{kkcon}
\eea 
where $\delta=\delta(\theta)$ takes varies between $\delta=16$
for $\theta =\pi/2$ to  $\delta \rightarrow \infty$ when $\theta
\rightarrow 0$. Note that the latter limit corresponds 
 to the conservation of KK momenta in the
absence of  localization as seen in Figure 2. The terms $\A^{cont}_{w}$ and 
$ \A^{cont}_{osc}$ contain  the contribution of long string streched between the intersections while winding around the compact dimension and the ones from
heavy string oscillation modes, respectively. In the large compactification 
radius limit $\A^{cont}_{w}$ is exponentially suppressed and we are left
with\footnote{The generic cases with finite radii can be found in 
\cite{ABL}}: 
\bea
\A^{cont}_{osc} = -\left[   {\bar \psi}^{(1)} \gamma_{M} \psi^{(2)} 
{\bar \psi}^{(4)} \gamma^{M} \psi^{(3)} \right ] \, \,  (\frac{g_s}{ M_s})^2 \,  \int_0^1 \frac {dx}{x} ( 
\frac {1}{ [F_\theta (x)]^2 } -1 )
\nonumber
\eea
where $\theta$ is the angle between the branes. 
For $\theta \rightarrow \frac{\pi}{2}$ we have $ \int_0^1 \frac {dx}{x} ( 
\frac {1}{ [F_\theta (x)]^2 } -1 )  \rightarrow 0.59$. 
For $\theta \rightarrow 0$, \,  $F_\theta (x)\rightarrow 1$
and this contact term vanishes. There is no tree level dimension six 
effective operator in the case of open strings ending on parallel branes
but the final amplitude can be written as:
\bea
\A (s,t) = \A_{point} (s,t) \cdot {\G(1-l_s^2 s) \G(1-l_s^2 t) \over \G(1-l_s^2s-l_s^2t)} = \A_{point} (s,t) \cdot \left[1 - {\pi^2\over 6}  {st\over M_S^4} + \cdots \right]
\label{summary1}
\eea
where $\A_{point}$ is the result usually derived from  the
(up-to-two-derivatives) low energy effective Lagrangian, while the
dimension-eight operator
here proportional to ${st\over M_S^4}$
represents the tree-level lowest order correction and  originates from the
form factor due to the string-like structure.

\section{ Experimental constraints on extra dimensions }\label{sec:exp}

\subsection{The scenario:}

In order to pursue further, we need to provide the quantum numbers and 
couplings of the relevant light states. We consider (see figure 3):
\begin{itemize}
\item Closed strings correpond to gravitons which describe fluctuations of the metric  propagate in the whole space. 

\item The gauge bosons propagate on a $(3+d)$-branes. They corresond on figure 3 to the open strings with both ends on the big brane.  

\item The matter fermions, quarks and leptons, are localized on 
3-branes (the small branes inside  bigger one on figure 3) and have no KK excitations.  
Our results strongly depend on this assumption. Instead, the possible localization of the Higgs scalar, as well as the possible
existence of supersymmetric partners do not lead to major modifications 
for most of the obtained bounds. 
\end{itemize}

\begin{figure}[t]
 \vspace{9.0cm}
\includegraphics{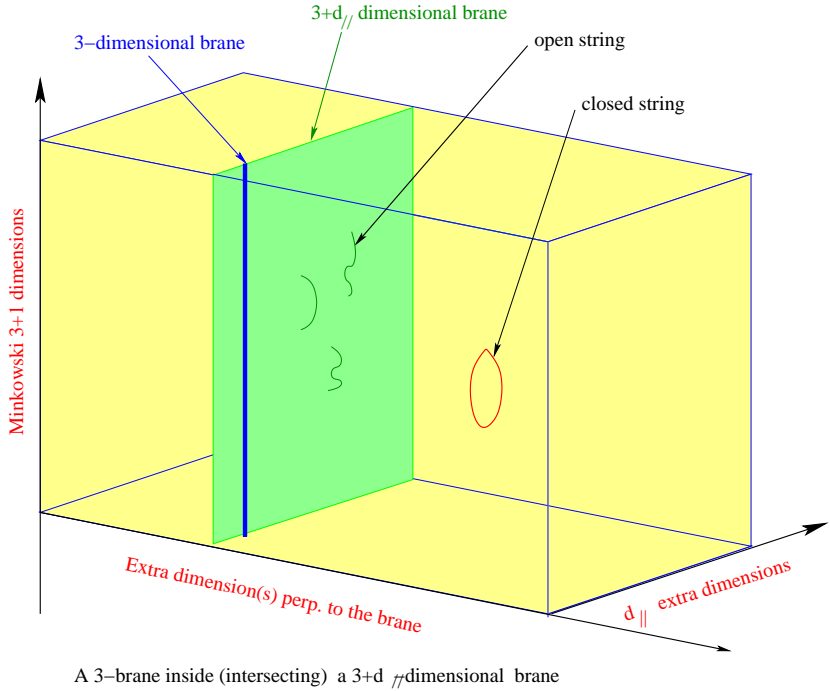}
 \caption{\it The geometrical set-up of our scenario for experimental bounds.
   \label{exfig3} }
\end{figure}


\subsection{Extra-dimensions along the world brane: KK excitations of gauge
bosons}

\begin{table}[t]
\caption{Limits on $R^{-1}_\parallel$ in TeV at present and future
colliders. The luminosity  is given in  fb$^{-1}$.\label{tab:para}}
\vspace{0.4cm}
\begin{center}
\begin{tabular}{  | c | c | c | c | l |} 
\hline
  & & & &
\\   Collider & Luminosity &  Gluons  & $W^{\pm}$ & $\gamma + Z$   \\ 
\hline\hline
\mco{5}{|c|}{Discovery of Resonances}   \\ \hline
  LHC     & 100        &  5  & 6  & 6               \\ \hline\hline
\mco{5}{|c|}{Observation of Deviation}   \\ \hline
 LEP 200    &$4\times 0.2$  & - & -  & 1.9 \\ \hline
 TevatronI & $0.11$ &  -  & - & 0.9  \\ \hline 
 TevatronII & 2 &  -  & - & $1.2$ \\ \hline 
  TevatronII  & 20 &  4 & - & 1.3 \\ \hline 
 LHC & 10& 15   & 8.2  & 6.7 \\ \hline
 LHC & 100& 20   & 14 &  12  \\ \hline
  NLC500 & 75& - & - & 8  \\ \hline
  NLC1000 & 200& - & - & 13  \\ \hline\hline
\end{tabular}
\end{center}
\end{table}

The experimental signatures of extra-dimensions are of two types:
\begin{itemize}

\item Observation of resonances due to KK excitations. This needs a collider
energy $\sqrt{s} \simgt 1/R_\parallel$ at LHC. The discovery limits in the
case of one extra-dimension are given in table 1.

\item Virtual exchange of the KK excitations which lead to measurable
deviations in cross-sections compared to the standard model prediction. The
exchange of KK states gives rise to an effective operator discussed above in section 3. For
$d>1$ the result depends then on both parameters $R_\parallel$ and
$M_s$. Example of analysis for $d=2$ can be found in Ref.~\cite{ABQ}. The
simpler case of $d=1$ has been studied in detail. Possible reaches
of colliders experiments~\cite{AAB,ABQ} are summarized in table 1.

\end{itemize}

Provided with good statistics, there are some ways to help distinguish the corresponding signals from other
possible origin of new physics, such as models with new gauge bosons: 
(i) the observation of resonances located
practically at the same mass value; (ii) the heights and widths of the
resonances are directly related to those of standard model gauge bosons in
the corresponding channels; (iii)  the size of virtual effects do not
reproduce a tail of Bright-Wigner resonance and a deep is expected just
before the resonance of the photon+$Z$, due to the interference between the
two.

\subsection{Extra-dimensions transverse to the brane world: KK excitations
of gravitons}\label{subsec:miss}

During a collision of center of mass energy $\sqrt{s}$, there are 
$(\sqrt{s}R_{\perp})^{d_\perp}$ KK excitations of gravitons with mass
$m_{KK\perp}<\sqrt{s}< M_s$, which can be emitted to the bulk. 
Each of these states 
looks from the four-dimensional point of view as a massive, quasi-stable, 
extremely weakly coupled ($s/M^2_{pl}$ suppressed) particle that escapes
from the detector. The total effect is a missing-energy cross section
roughly of order 
$\frac {(\sqrt{s}R_{\perp })^n} {M^2_{pl}} \sim \frac{1}{s} 
{(\frac{\sqrt{s}}{M_s})^{n+2}}$. 
Explicit computation of these effects leads to the bounds given in table
2~\cite{missing} while astrophysical bounds~\cite{astcos,supernovae} arise 
from the
requirement that the radiation of gravitons should not carry on too much
of the gravitational binding energy released during core collapse of
supernovae. 
The best cosmological bound~\cite{COMPTEL} is obtained from requiring that
decay of bulk gravitons to photons do not generate a spike in the energy
spectrum of the photon background measured by the COMPTEL instrument. The
bulk gravitons  are themselves expected  to be produced just before
nucleosynthesis due to thermal radiation from the brane. The limits assume
that the temperature was at most 1 MeV as nucleosynthesis begins, and become
stronger if this temperature is increased. While the obtained bounds for $R_\perp^{-1}$ are 
smaller than those that could be checked in table-top experiments probing 
macroscopic gravity at small distances, one should keep in mind that 
larger radii are allowed if one relaxes the assumption of isotropy, 
by taking for instance two large dimensions with different radii.

\begin{table}[t]
\caption{Limits on $R_\perp$ in mm from missing-energy
processes.\label{tab:exp3}}
\vspace{0.4cm}
\begin{center}
\begin{tabular}{  | c | c | c | l |} 
\hline
  & & &
\\   Experiment & $R_\perp (n=2)$ & $R_\perp (n=4)$ & $R_\perp (n=6)$ \\ 
\hline\hline
\mco{4}{|c|}{Collider bounds}   \\ \hline

 LEP 2   & $4.8\times 10^{-1}$ & $1.9\times 10^{-8}$  & 
                              $6.8 \times 10^{-11}$ \\ \hline
  Tevatron  &   $5.5 \times 10^{-1}$  & $1.4 \times 10^{-8}$ 
              & $4.1 \times 10^{-11}$ \\ \hline 
  LHC &  $4.5 \times 10^{-3}$   & $5.6\times 10^{-10}$  & 
                              $2.7 \times 10^{-12}$  \\ \hline
  NLC & $1.2\times 10^{-2}$  & $1.2\times 10^{-9}$  & 
                              $6.5 \times 10^{-12}$  \\ \hline\hline
\mco{4}{|c|}{Present non-collider bounds}   \\ \hline
  
SN1987A   &  $3 \times 10^{-4}$   & 
           $1 \times 10^{-8}$ 
                 & $6 \times 10^{-10} $ \\ \hline
COMPTEL &  $5 \times 10^{-5}$   & - & 
                              - \\ \hline
\end{tabular}
\end{center}
\end{table}

\section{Dimension-Six Effective Operators: }

 The dimension-six effective operators are generically
parametrized as \cite{fourferm}:
\bea
\L_{eff} = \frac{4 \pi}{(1+\varepsilon )\Lambda^2} \sum_{a,b=L,R}\eta_{ab} 
{\bar \psi_a} \gamma^\mu \psi_a {\bar {\psi'_b}} \gamma_\mu \psi'_b
\label{Lambda}
\eea 
with $\varepsilon=1$ (0) for $\psi=\psi'$ ($\psi \neq \psi'$),
where $\psi_a$ and $\psi'_b$  are left ($L$) or right ($R$) handed
spinors. $\Lambda$ is the scale of contact interactions  and
$\eta_{ab}$ parametrize the relative strengths of various helicity
combinations. The generic analysis of these operators can be found in 
\cite{ABL}. We summarize here some of the results.

For $\psi\neq\psi'$  the contributions only from the exchange of the
massive open string states on the small brane lead to 
 parameters in eq. (\ref{Lambda}) as: 
\bea 
\eta_{LL}= \eta_{RR}=\eta_{LR}=\eta_{RL}= 1\, , \, \, \, \, \, \, 
\Lambda \simeq \sqrt{\frac{4\pi}{{0.59} g_s}} M_s
\label{fres1}
\eea
The signs and relative ratios of the different terms in
(\ref{Lambda}) correspond to what is usually refered to as
$\Lambda_{VV}^{+}$. The present bounds from LEP \cite{LEP} are of  the
order of $\Lambda_{VV}^{+} \simgt 16$ TeV which for $g_s=g_{YM}^2 \sim
1/{2}$, with $g_{YM}$ the gauge coupling, leads to  $M_s \simgt 2.5$
TeV. A stronger bound can be obtained from the analysis
of high precision low energy data in the presence of effective
four-fermion operators that modify the $\mu$-decay amplitude. 
 Using the results of ref.~\cite{strumia}, we
obtain $M_s \simgt 3.1$ TeV. 

In the case $\psi = \psi'$ as for Bhabha scattering in $e^+ e^-$
there is an  additional contribution to the effective operator coming
from the operators that are  associated with the exchange of other massives
oscillation modes  leading instead to $0.75\, \eta_{LL}=0.75\, \eta_{RR}\simeq\eta_{LR}=\eta_{RL}= 1$.

On the other hand, the  contact interactions due to exchange of KK excitations
give rise (for $d_\parallel = 1$ to \cite{AB}: 
\bea 
\L_{eff}^{KK}
\simeq -\frac{ \pi^2}{3(1+\varepsilon )} R^2 g_s
\sum_{a,b=L,R}\eta_{ab}  {\bar \psi_a} \gamma^\mu \psi_a {\bar
{\psi'_b}} \gamma_\mu \psi'_b
\label{Lambda2}
\eea 
Experimental constraints on such operators translate into lower
bounds on the scale of compactification. For instance exchanges
of KK excitations of photon corresponds  to $\eta_{ab} = 1$ and 
 $g_s/4\pi = 1/128$ from which we obtain a bound  $R^{-1} \simgt 2.2$ TeV, 
using  LEP bounds \cite{LEP} $\Lambda_{VV}^{-} \simgt 14$ TeV.
Low energy precision electroweak data lead instead to 
$R^{-1} \simgt 3.5$ TeV \cite{KK2}.
\begin{figure}[t]
 \vspace{9.0cm}
\includegraphics{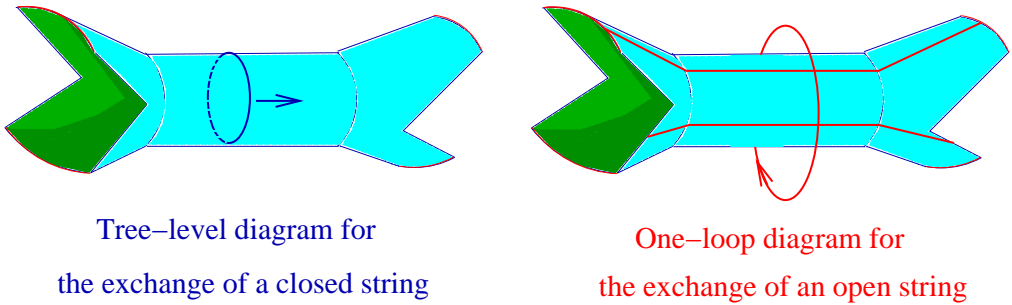}
 \caption{\it The exchange of virtual gravitons. \label{exfig4} }
\end{figure}

\subsection{Dimension-Eight Effective Operators:}

We consider two
generic sources for dimension-eight operators: (i) Form factors due to the 
extended nature of strings eq.~(\ref{summary1})(ii) exchange of virtual KK
excitations of bulk fields (gravitons,...).

The limit obtained
from   dimension-eight operators (i) is of order
  $M_s \simgt 0.63$ TeV \cite{Peskin,lim}. Instead (ii) can not provide
  reliable model dependent results. 
The exchange of virtual KK excitations of bulk gravitons is described
in the effective  field theory by an amplitude involving the sum
$\frac {1}{M_p^2}\sum_n \frac {1}{s-\frac{{\vec n}^2}{R_\perp^2}}$. For
$n > 1$, this sum diverges. This means it is sensitive to the UV cut-off
physics thus cannot be compute reliably  in field theory.
In  string models it reflects the 
ultraviolet behavior of open string one-loop diagrams which are  
compactification dependent.

In order to understand better this issue, it is important to remember that
 gravitons and other bulk particles correspond to
excitations of closed strings. Their tree-level exchange of order $g_s^2$ 
is described  by a  cylinder which  can also be seen  as an annulus
corresponding to an open string  describing a loop  (see figure 4). 
First,  the result of such one-loop diagrams 
are  compactification dependent. Second, they correspond to
 box diagrams in a gauge theory
which are  of order $g_{YM}^4$ thus samller by a factor 
$g_s = g_{YM}^2$ compare to the ones in (i).

\section*{Acknowledgments}
This work was supported in part by the European Commission under TMR
contract ERBFMRX-CT96-0090 and RTN contract HPRN-CT-2000-00148, and in
part by the INTAS contract 99-0590. The work of K.B. was also partially
supported by NSERC of Canada and  Fonds FCAR du Qu\'ebec.

\end{document}